# Doped and non-doped organic light-emitting diodes based on a yellow carbazole emitter into a blue-emitting matrix.

H. Choukri[a], A. Fischer[a], S. Forget[a*], S. Chenais[a], M-C. Castex[a], B. Geffroy[b], D. Adès[c], A. Siove[c]

[a] *Laboratoire de Physique des Lasers, CNRS/Université Paris 13, 93430 Villetaneuse, France*
[b] *Laboratoire des Composants Hybrides, DTNM/LITEN/CEA-Saclay, 91191 Gif sur Yvette Cedex, France*
[c] *Laboratoire des Polymères de Spécialité, CNRS/Universités Paris 13, 93430 Villetaneuse, France*

**Abstract**

A new carbazole derivative with a 3,3'-bicarbazyl core 6,6'-substituted by dicyanovinylene groups (6,6'-*bis*(1-(2,2'-dicyano)vinyl)-*N,N'*-dioctyl-3,3'-bicarbazyl; named (OcCz2CN)$_2$, was synthesized by carbonyl-methylene Knovenagel condensation, characterized and used as a component of multilayer organic light-emitting diodes (OLEDs). Due to its π-donor–acceptor type structure, (OcCz2CN)$_2$ was found to emit a yellow light at $\lambda_{max}$=590 nm (with the CIE coordinates x=0.51; y = 0.47) and was used either as a dopant or as an ultra-thin layer in a blue-emitting matrix of 4,4'-bis(2,2'-diphenylvinyl)-1,1'-biphenyl (DPVBi). DPVBi (OcCz2CN)$_2$-doped structure exhibited, at doping ratio of 1.5 weight %, a yellowish-green light with the CIE coordinates ($x = 0.31$; $y = 0.51$), an electroluminescence efficiency $\eta_{EL}$=1.3 cd/A, an external quantum efficiency $\eta_{ext}$= 0.4 % and a luminance $L$= 127 cd/m$^2$ (at 10 mA/cm$^2$) whereas for non-doped devices utilizing the carbazolic fluorophore as a thin neat layer, a warm white with CIE coordinates (x = 0.40; y= 0.43), $\eta_{EL}$= 2.0 cd/A, $\eta_{ext}$= 0.7 %, $L$ = 197 cd/m$^2$ (at 10 mA/cm$^2$) and a color rendering index (CRI) of 74, were obtained. Electroluminescence performances of both the doped and non-doped devices were compared with those obtained with 5,6,11,12-tetraphenylnaphtacene (rubrene) taken as a reference of highly efficient yellow emitter.

*Keywords:* carbazole; OLED; doped and non-doped devices; white emission

## 1. Introduction

Organic light emitting diodes (OLEDs) have attracted great attention for full-colour flat-panel displays since the demonstration of efficient electroluminescent devices [1]. Moreover, the tremendous progress in device luminous efficiency [2] and the emerging viability as a commercial display open new markets for white OLEDs (WOLEDs), like backlight for liquid crystal displays or lighting sources. WOLEDs can be considered as potential large area next generation of solid state lighting sources to replace traditional incandescent white light sources, thanks to their potential of energy saving, their high efficiencies and their possibility to fabricate thin and flexible devices [3]. To achieve white emission, various methods have been used such as *e.g.* multilayer diodes using the combination of the three primary colours Red, Green and Blue (RGB), multiple-quantum-well architecture [4], or exciplex formation [5]. Among these various devices, numerous doped-type WOLEDs using two emission colours to produce white have been fabricated [6,7] but it remains technologically difficult to attain an accurate control of the dopant concentration at low doping ratio through the co-evaporation process. We recently utilized the delta-doping technique [8], a way to colour-control, including balanced white emission, in a multilayer non-doped OLED [9-11] based on blue matrices, in which we inserted an ultrathin layer of rubrene as a yellow emitter [12]. Hence, we intended to test another yellow material with a different structure and coming from the carbazolic molecular derivatives.

Indeed, these materials have been widely used in optoelectronics [13] and particularly as emitters in OLEDs by several of us [14-17]. The main goal of our work is turned towards synthetic chemistry of luminescent organic materials belonging to the carbazole family. Unsusbstituted carbazolic members are well-known for their good hole transporting properties and for their blue-light emitting capabilities (see *e.g.* review ref. [14]).

With such a goal, we report, herein, on the synthesis of a novel carbazole derivative with an electron-donating 3,3'-bicarbazyl core linked, at the 6,6'-positions, to electron-accepting dicyanovinylene groups *i.e.* 6,6'-*bis*(1-(2,2'-dicyano)vinyl)-*N,N'*-dioctyl-3,3'-bicarbazyl (named (OcCz2CN)$_2$). This compound, was synthesized by carbonyl-methylene Knovenagel condensation between bicarbazyl dialdehyde and dicyanomethane and was characterized by $^1$H and $^{13}$C NMR, electronic absorption and emission spectroscopies and cyclic voltammetry. Molecular structure of (OcCz2CN)$_2$, is shown in Figure 1. Due to its π-donor-acceptor type structure, (OcCz2CN)$_2$ exhibits an energy band-gap of $Eg$ = 2.4 eV and was found to emit a yellow light with Commission Internationale de l'Eclairage (CIE) coordinates of (x=0.51; y = 0.47). To the best of our knowledge, it is the first carbazolic derivative exhibiting such properties.

It was used in multilayer OLEDs either, as a dopant of a blue-emitting matrix of 4,4'-bis(2,2'-diphenylvinyl)-1,1'-biphenyl (DPVBi) or as an ultrathin layer inserted in the latter. For a doping ratio of 1.5 weight %, devices exhibited modest electroluminescence (EL) performances *e.g.* an electroluminescence efficiency $\eta_{EL}$=1.3 cd/A, an external quantum efficiency $\eta_{ext}$= 0.4 % and a luminance $L$= 127 cd/m$^2$ (at 10 mA/cm$^2$), a yellowish-green light being emitted. For the non-doped diodes utilizing a thin layer of (OcCz2CN)$_2$ within the exciton diffusion zone in DPVBi, diodes exhibited better EL performances and an emission colour varying from yellow to white depending on the position of the carbazole film within the DPVBi layer.

---

* Corresponding author

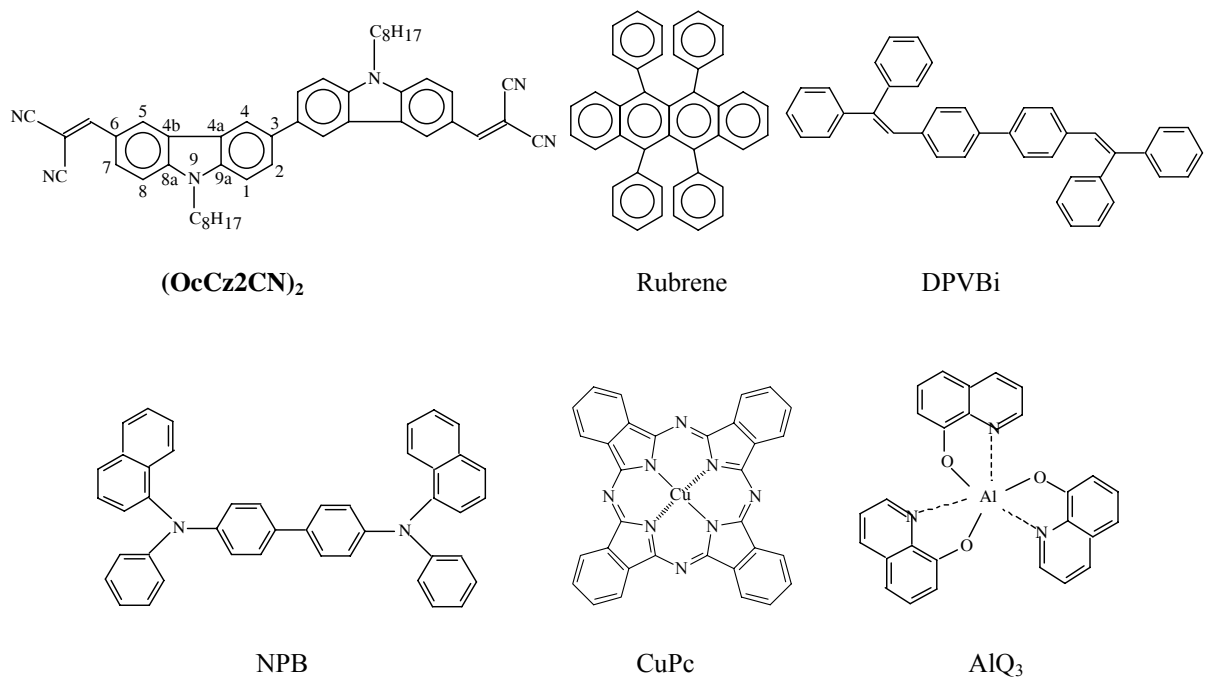

Figure 1. Molecular structures of the organics used for doped and non-doped OLEDs with (OcCz2CN)$_2$ as a yellow fluorophore in blue-emitting DPVBi layer.

For such delta-doped devices utilizing a carbazolic thin neat layer, white emission with CIE coordinates (x = 0.40; y= 0.43), $\eta_{EL}$= 2.0 cd/A, $\eta_{ext}$= 0.7 %, $L$ = 197 cd/m$^2$, were obtained. To the best of our knowledge, it is the first report of a white OLED using a carbazolic material as a thin emitting layer. EL performances of both devices were compared with those obtained with rubrene which is by far the best organic yellow fluorophore and thereby which is taken as a yellow quantum efficiency reference in the literature and is known to enhance multilayer device efficiency and stability [18,19].

## 2. Experimental

### 2.1. Materials

Aluminium tris(8-hydroxyquinolinate) (AlQ$_3$), copper pthalocyanine (CuPc), 4,4'-bis[*N*-(1-naphtyl)-*N*-phenylamino]biphenyl, (NPB), 4,4'-bis(2,2'-diphenyl vinyl)-1,1'-biphenyl (DPVBi), 5,6,11,12-tetraphenylnaphtacene (rubrene) were purchased from H.W. Sands Company. Molecular structures are shown in Figure 1.

The synthetic route of (OcCz2CN)$_2$ is shown in Figure 2. All the solvents were purified by the usual methods prior to utilization. (OcCz2CN)$_2$ was synthesized by carbonyl-methylene Knovenagel condensation between *N,N'*-dioctyl-3,3'-bicarbazyl-6,6'-dicarbaldehyde and dicyanomethane as the active methylene compound. *N,N'*-dioctyl-3,3'-bicarbazyl-6,6'-dicarbaldehyde was prepared according to the method we previously described [20]. We briefly recall, that the dialdehyde was obtained by diformylation of *N,N'*-dioctyl-3,3'-bicarbazyl (itself prepared by oxidative dimerization of *N*-octyl-carbazole by iron trichloride in chloroform) by *N,N*-dimethylformamide (DMF) in *o*-dichlorobenzene (ODCB) (first reaction step of Figure 2). After column chromatography with chloroform/hexane (3:1) and recrystallization from diethylether, yellow crystals of the dialdehyde were obtained in 55% yield.

Knovenagel condensation is carried out as follows. Equimolar quantities of the dialdehyde and dicyanomethane (2 mmol), were dissolved in 20 ml of dry tetrahydrofuran (THF) and introduced into a three-necked flask fitted with a condenser and a nitrogen inlet. A 2 mmol solution of tetrabutylammonium hydroxide (Fluka) in an isopropanol-methanol mixture (10:1), was then added to the reaction medium under nitrogen. After refluxing for 1 hour, reaction was stopped by cooling the reaction mixture at 0°C, and poured into ice-water (150 ml). The resulting brownish solid obtained was filtered off and purified by column chromatography with chloroform/hexane (3:1). Orange powder was obtained in 72 % yield.

Anal. Calc. for $C_{48}H_{48}N_6$: C 81.31, H 6.82, N 11.87; Found: C 81.79, H 6.65, N 11.53. $^1$H NMR (CDCl$_3$, 200MHz, δ ppm) : 8.4 ($H_{5,5'}$), 8.1 ($H_{4,4'}$), 7.8 ($H_{2,2'}$), 7.6 ($H_{7,7'}$), 7.5 (2 H vinylene), 7.3 ($H_{1,1'}$), 7.2 ($H_{8,8'}$), 4.1-1.1 (28H CH$_2$), 0.7 (6H CH$_3$). $^{13}$C NMR (CDCl$_3$, 50MHz, δ ppm) : 159.2 (CH-CN vinylene), 143.8 ($C_{8a,8'a}$), 140.1 ($C_{9a,9'a}$), 134.3 ($C_{3,3'}$), 128.3 ($C_{5,5'}$), 126.6 ($C_{7,7'}$), 124.4 ($C_{2,2'}$), 123.4 ($C_{6,6'}$), 122.9 ($C_{4b,4b'}$), 122.1 ($C_{4a,4a'}$), 118.9 ($C_{4,4'}$), 114.5 and 113.6 (CN(s)

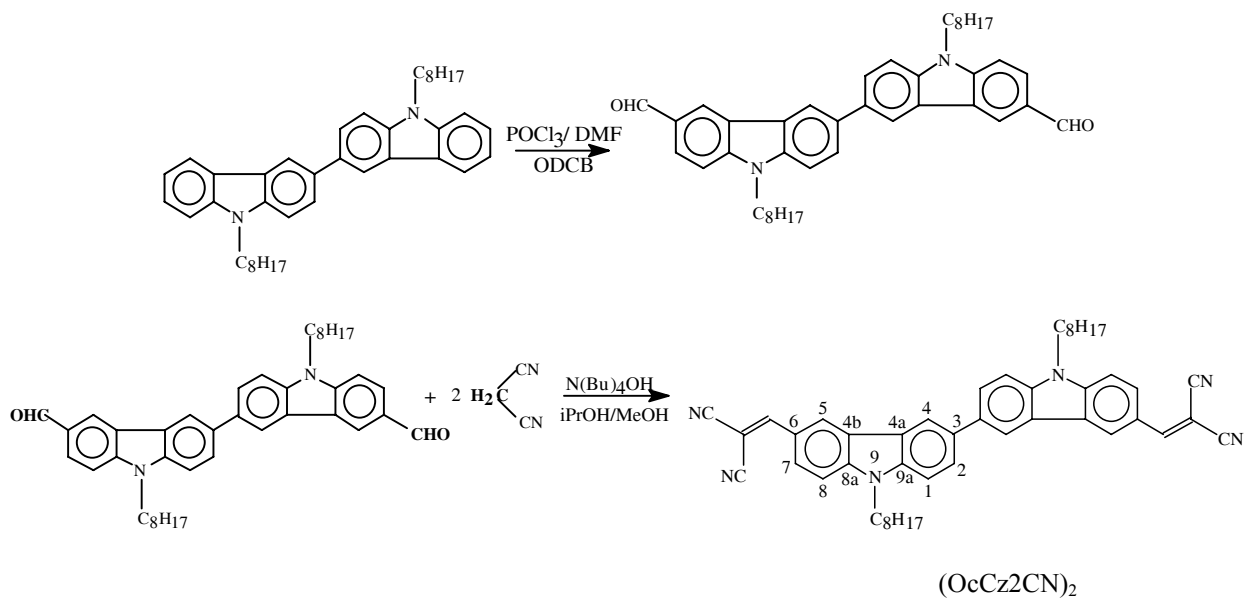

Figure 2. Synthetic route of bicarbazyldialdehyde and (OcCz2CN)$_2$

cis, trans), 109.6 ($C_{8,8'}$), 109.3 ($C_{1,1'}$), 76.3 (C-CN vinylene), 43.5-22.2 ($CH_2$ (s) octyl), 13.8 ($CH_3$).

*2.2. Physico-chemical characterization*

$^1$H and $^{13}$C NMR spectra were recorded in solution in $CDCl_3$ with a Brucker AC200 spectrometer at 200 and 50 MHz respectively; all chemical shifts ($\delta$ in ppm) were referenced to tetramethylsilane (TMS). UV-visible spectroscopy was carried out using a Perkin Elmer spectrometer (model Lambda 12) with 0.5 cm path quartz cells. For solid-state measurements, $(OcCz2CN)_2$ chloroform solutions were cast onto quartz plates. Optical band gap $E_g$ was estimated from the absorption threshold of the spectra. Photoluminescence spectra were obtained with a Safas model flx-Xenius spectrofluorometer with a Xenon 150 W lamp. Cyclic voltammetry (cv) experiments of solution-cast films of $(OcCz2CN)_2$ onto conducting support were carried out in acetonitrile (HPLC grade, Fisher Scientific) containing 0.3 M of $LiClO_4$ (Aldrich) as a supporting electrolyte in a one-compartment cell equipped with three electrodes. The working anode was glassy carbon, the counter-electrode a platinum wire, the reference electrode being a saturated calomel electrode (SCE) at 4.75 eV.

*2.3. OLEDs fabrication and characterization*

Figure 3 details the layer structures of the two types of devices with the detail of the organic compounds and the thickness of each layer of the two types of devices a)-type (DPVBi layer doped with yellow emitter) and b)-type (non-doped type, with a thin layer of yellow emitter in DPVBi). Indium tin oxide (ITO)-coated glass with sheet resistance of about 15 Ω/□ was obtained from Asahi. Prior using, the glass was cleaned by sonication in a detergent solution and then rinsed in de-ionized water. The organic compounds are deposited onto the ITO anode by sublimation under high vacuum ($10^{-7}$ Torr) with a rate of 0.2 – 0.3 nm/s. An in-situ quartz crystal was used to monitor the thickness of the vacuum depositions. The active area of each OLED was 0.3 $cm^2$. In the doped configuration, we sequentially deposited onto the ITO anode: a thin (10 nm) layer of copper phtalocyanine (CuPc) as a hole-injecting layer, then a 50 nm thick NPB layer as a hole-transporting layer (HTL). A 60 nm thick DPVBi layer doped with the yellow emitter $(OcCz2CN)_2$ is then deposited and acts as a blue matrix and an electron-transporting layer (ETL). The doping ratio is controlled by co-evaporation of host and dopant and was optimized to be 1.5% (in weight percent) of $(OcCz2CN)_2$, the accuracy of the doping rate being around 10%. Finally, a thin (10 nm) $AlQ_3$ layer was used as an electron-injecting layer onto the top cathode consisting in 1.2 nm of LiF capped with 100nm of Al. In the non-doped configuration, the same sequential process was followed, keeping constant the thicknesses of the organic layers, the only difference from the doped configuration being that an ultrathin layer of $(OcCz2CN)_2$ was inserted within the DPVBI layer, at a position $d$ from the NPB/DPVBi interface in the range $0 \leq d \leq 10$ nm. Such a range corresponds to the exciton diffusion domain within the DPVBi side [12]. Electroluminescence spectra and chromaticity coordinates were recorded with a PR 650 SpectraScan spectrophotometer at a constant current density of 10mA/$cm^2$. The Current-Voltage and Current-Luminance (I-V and I-L) characteristics of the diodes were measured with a regulated power supply (ACT 10 Fontaine) combined with a multimeter and a 1 $cm^2$ area silicon calibrated photodiode (Hamamatsu). All the measurements were performed at room temperature and under ambient atmosphere, without any encapsulation. An important quality of white light emitters for illumination purpose is the Color

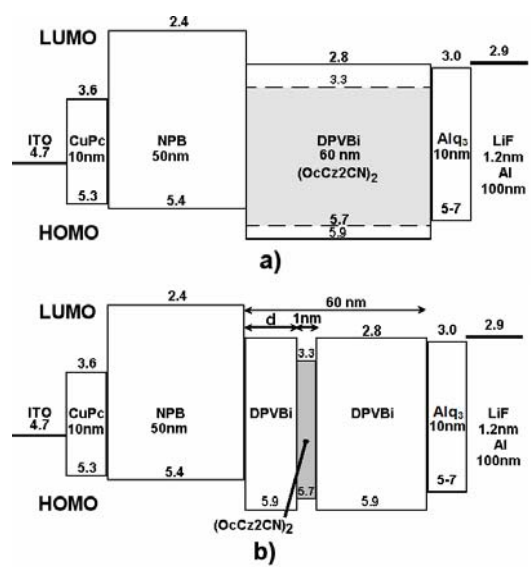

Figure 3. Configuration of the multilayer OLEDs; (a): $(OcCz2CN)_2$-doped DPVBi, (b): non-doped, with a 1 nm thick $(OcCz2CN)_2$ film inserted in DPVBi layer. HOMO level of $(OcCz2CN)_2$ was obtained from cyclic voltammetry. LUMO level was deduced from the optical energy band-gap $E$g and HOMO values.

Rendering Index (CRI). The CRI defines how well colors are rendered under different illumination conditions in comparaison to a standard (*i.e.* a thermal radiator or daylight). As defined by the Commission Internationale de l'Eclairage [21], a set of eight calibrated samples with different absorption spectra and a reference illuminant are used to define the CRI. The reference source used for the CRI calculation is not given in the CIE definition. To choose the appropriate illuminant, we first calculate the Correlated Color Temperature (CCT) (according to reference [22]) of our White OLED. The reference source used for the CRI calculation is then the black body at this temperature if the CCT is less than 5500K, and the D65 Daylight illuminant otherwise.

## 3. Results and discussion

*3.1. Synthesis, characterization and physico-chemical properties*

As illustrated in Figure 2 the synthesis of $(OcCz2CN)_2$ proceeds in two steps. At first,

bicarbazyldialdehyde was prepared by formylation of bicarbazyl according to the method that we previously described [20]. In a second step, (OcCz2CN)$_2$ was obtained by Knovenagel condensation of the bicarbazyldialdehyde as the carbonyl reactive compound, with dicyanomethane as the methylene active derivative. Both the $^1$H and $^{13}$C NMR spectra and elemental analysis are in agreement with the chemical structure of (OcCz2CN)$_2$ shown in Figure 1.

Absorption and photoluminescence spectra of 50nm thin film of the tetracyanobicarbazyl derivative deposited onto a quartz plate are shown in Figure 4. Absorption spectrum is obtained by exciting the thin film in the range of 300–700 nm wavelengths. The photoluminescence spectrum is so determined by exciting the film around the absorption maxima.

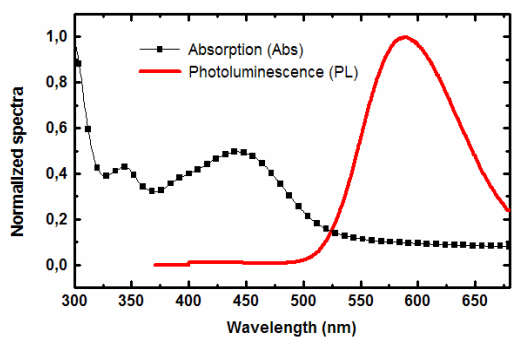

Figure 4. Absorption (line + square) and photoluminescence (solid line) spectra of the 50nm thick film of (OcCz2CN)$_2$.

As already reported for several donor-acceptor carbazole-based molecules, substitution in 3-position of the electron-donor heterocycle by cyanovinylene electron-withdrawing groups leads to an internal charge transfer which shows up in the visible domain of the optical spectrum [23].
In the case of (OcCz2CN)$_2$, *i.e* 3,3'-dimer of OcCz2CN, the charge-transfer band is peaking at $\lambda_{max}$= 450 nm. From the onset of the absorption threshold of the spectrum the energy gap value $E$g was found to be close to 2.4 eV. Furthermore, it is noteworthy that π-conjugation within the bicarbazyl core provokes, as expected, a red-shift of the visible band, since the corresponding signal appears at $\lambda_{max}$= 428 nm for films of OcCz2CN monomer ($E$g=2.7 eV) [23]. Hence, a yellow-light emission is observed for (OcCz2CN)$_2$ with a maximum at $\lambda_{max}$= 590 nm (Figure 4) whereas a green one is observed for the monomer derivative ($\lambda_{max}$= 528 nm). Such an effect of the extension of π-conjugation on the shift of the emission color appeared in red-emitting *N*-arylcarbazoles substituted in 3,(6)-position(s) by three conjugated double bonds ended by a dicyano group, recently reported by Fu *et al.* [24].

Cyclic voltammetry of (OcCz2CN)$_2$ (at a scan rate of 10 mV/s), allowed an estimation of its ionization potential IP (HOMO level). An oxidation process at $E_{ox}$ = 0.95 V/SCE (HOMO level at 5.7eV) is observed corresponding to the formation of carbazolic radical-cations [25]. This value is 0.10 V higher than that observed for 3,3' unsubstituted bicarbazyl, as expected from the effect of the electron-acceptor groups. Indeed, it has been shown that such substituents make the carbazole core less oxidizable and lead to an anodic shift of the oxidation potential. On the opposite, introduction of acceptor groups is expected to increase the electronic affinity (EA). Nevertheless, no clear reduction signal was observed in the electrochemical reduction domain of cv and LUMO level value was deduced from the optical energy band-gap $E$g and HOMO values according to the expression $E_{HOMO} - E_{LUMO} = E g$. Such estimation leads to a value of 3.3 eV for the LUMO.

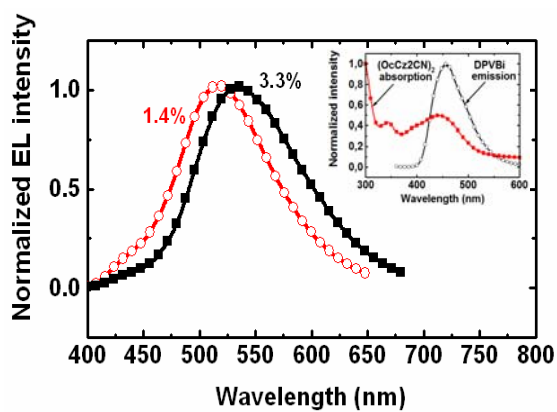

Figure 5. Electroluminescence spectra at a current density of 10mA/cm² of doped OLEDs with 1.4% and 3.3% wt of (OcCz2CN)$_2$ in DPVBi. Inset: proposed (OcCz2CN)$_2$ absorption and DPVBi emission spectra.

3.2. *Electroluminescence*

Doped and non-doped OLED structures are depicted Figure 3.

*Doped-OLEDs*

EL spectrum of doped devices in configuration (a) *i.e.* with the following structure:
ITO/CuPc/NPB/DPVBi:(OcCz2CN)$_2$/AlQ$_3$/LiF/Al
(DPVBi doped with 1.5% and 3.3% of the yellow emitter is shown on Figure 5).

Although NPB is also a blue fluorescent emitter, the position of the EL blue luminescence peak in our devices match perfectly (after small correction due to calculated microcavity effects) the DPVBi PL peak, which was measured on a separate substrate and differs from about 20 nm from the NPB one. We consequently consider that the blue emission is due to DPVBi in our devices.
A good overlapping between (OcCz2CN)$_2$ absorption and DPVBi emission (inset Figure 5) leads to an efficient Förster transfer process. With increased (OcCz2CN)$_2$ concentration from 1.4 to 3.3 wt %, the wavelength of the peak red shifts from 520 to 540nm. This red-shift of the EL peak could be attributed to polarization phenomena [26]. Meanwhile, external quantum efficiency $\eta_{ext}$ decreases from 0.4 to 0.3 %.

Such an effect of the dopant concentration on $\eta_{ext}$ may arise from a quenching process. EL parameters of doped devices are presented in Table 1.

Doping DPVBi with 1.5% wt of rubrene, taken as a reference of highly efficient yellow emitter, led to an important improvement of all the EL parameters as e.g. a $\eta_{ext}$ = 2.5 % (Table 1), a greenish-yellow light being observed. These results show that rubrene is a much more efficient dopant than $(OcCzCN_2)_2$ although having close LUMO and HOMO levels (3.2 eV against 3.3 eV and 5.4 eV against 5.7 eV respectively). This is not surprising as it has been reported that rubrene dispersed in a host layer acts as a very efficient trap charge carrier and recombination center [27]. In order to attain a color-control we tested another type of OLED configuration i.e. a non-doped device with a thin film of $(OcCz2CN)_2$ inserted in the DPVBi layer (Figure 3b).

*Non-doped-OLEDs*

A 1 nm-thick layer of the carbazolic dimer was inserted within the DPVBi blue-emitting matrix at a position $d$ from the NPB/DPVBi interface in the range $0 \leq d \leq 10$ nm. By using an ultrathin rubrene sensing layer we recently found for the same structure that such a $d$ range corresponds to the exciton diffusion domain within the DPVBi layer (diffusion length around 9nm) [12].

Typical emission spectra for different values of $d$ (5, 7 and 10 nm) are shown on Figure 6 with corresponding CIE coordinates in Inset. In these configurations, the emission spectrum consists in two blue and yellow peaks. The relative intensity of these peaks is sensitive to the position of the carbazolic layer from the interface. With increased $d$ from 5 to 10nm, the amount of blue is increased by the exciton diffusion in the DPVBi matrix.

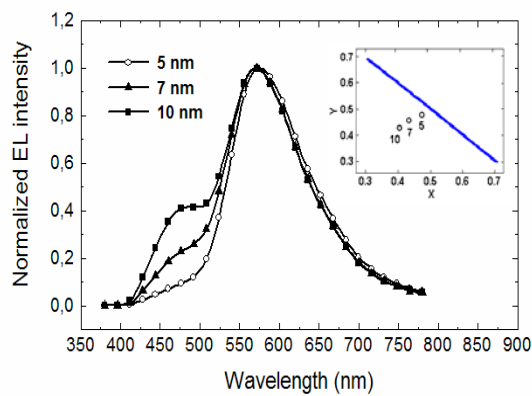

Figure 6. EL spectra for different positions $d$ of a 1nm thick layer of $(OcCz2CN)_2$ from the NPB/DPVBi interface. The spectra are normalized to the yellow peak at 568nm. Inset: corresponding CIE diagram (the solid line corresponds to the edge of the spectrum locus).

Influence of the position $d$ of the $(OcCz2CN)_2$ layer on performances and CIE coordinates for several devices are shown on Table 2. In absence of yellow emitter, a deep blue luminescence with CIE coordinates (0.17; 0.15), a quantum efficiency $\eta_{ext}$= 2.8 % typical for DPVBi were observed (see EL performances of undoped DPVBi Table 1). Pure yellow emission with CIE coordinates (0.51; 0.47) corresponding to the emission light of $(OcCz2CN)_2$ molecules is achieved when the carbazolic compound lies exactly at the interface ($d = 0$). Here, the energy transfer between DPVBi excitons and $(OcCz2CN)_2$ is complete, as a consequence of the efficient Förster process. Nevertheless, the quantum efficiency is $\eta_{ext}$= 0.3% around one order of magnitude lower than that of the blue OLED, showing that important quenching limits the efficiency. A warm white standard emission closed to – illuminant A- with CIE coordinates (0.40; 0.43), a $\eta_{ext}$ = 0.7 %, a $\eta_{EL}$ = 0.7 cd/A, a luminance $L$ = 932 cd/m$^2$ at 60mA/cm$^2$ (197 cd/m$^2$ at 10mA/cm$^2$) and a quite good Color Rendering Index of 74, were obtained for a position $d = 10$ nm. To the best of our knowledge, it is the first report of a white OLED using a carbazolic molecular compound as a thin emitting layer. Intermediary yellowish-green light (CIE: 0.47; 0.48) and yellow (CIE: 0.43; 0.46) lights were obtained for $d = 5$ nm and $d = 7$ nm respectively.

For comparison, devices utilizing 1nm of rubrene instead of $(OcCz2CN)_2$ have found to produce white light with similar CIE coordinates [12] and a quantum efficiency $\eta_{ext}$= 2.6 %, at a closier $d$ position from the emitting interface i.e. $d = 5$ nm from the emitting interface. We can also observe that when increasing $d$ from 0 to 10nm, CIE coordinates of rubrene-doped DPVBi decrease from yellow (0.51; 0.48) to blue-white (0.25; 0.23). In the case of the $(OcCz2CN)_2$, the decrease is weaker, from (0.51; 0.47) to (0.40; 0.43). At $d = 10$nm, blue emission is more important in the rubrene configuration.

**4. Conclusion**

A new carbazole derivative 6,6'-*bis*(1-(2,2'-dicyano)vinyl)-*N,N*'-dioctyl-3,3'-bicarbazyl (($OcCz2CN)_2$), was synthetised by Knovenagel condensation and used as an active molecule in multilayer organic light-emitting diodes (OLEDs). Due to the $\pi$-donor–acceptor conjugation developed between the electron-donor carbazolic moieties and the cyanovinylene electron-withdrawing groups, $(OcCz2CN)_2$ was found to emit a yellow light at $\lambda_{max}$=590 nm with CIE coordinates (x = 0.51; y = 0.47) and was used either as a dopant or as an ultra-thin layer in a blue-emitting matrix of 4,4'-bis(2,2'-diphenylvinyl)-1,1'-biphenyl (DPVBi). DPVBi $(OcCz2CN)_2$-doped structure exhibited, at doping ratio of 1.5 weight %, a yellowish-green light with CIE coordinates (x = 0.31; y = 0.51), an electroluminescence efficiency $\eta_{EL}$=1.3 cd/A, an external quantum efficiency $\eta_{ext}$ = 0.4 % and a luminance $L$= 127 cd/m$^2$ (at 10 mA/cm$^2$) whereas for non-doped devices utilizing a thin carbazole yellow-emitting layer inserted in a blue-emitting DPVBi matrix, white OLEDs carbazole-based have been achieved for the first time. A warm white light with CIE coordinates (x

= 0.40; y= 0.43), $\eta_{EL}$= 2.0 cd/A, $\eta_{ext}$= 0.7 %, $L$ = 197 cd/m$^2$ (at 10 mA/cm$^2$) and a quite good Color Rendering index (CRI) of 74, were obtained. Comparison between EL performances of devices utilizing our carbazole molecule with those based on the highly efficient rubrene reference show that the former are only 4 times lower. Increasing the quantum efficiency of our yellow light-emitting carbazoles would imply theoretical investigations coupled with subsequent experimental chemical engineering, which is beyond the scope of this paper.

## References

[1] C. W. Tang, S. A. Vanslyke, Appl. Phys. Lett. 51 (1987) 913.

[2] M. Ikai, S. Tokito, Y. Sukamoto, T. Suzuki, Y. Taga, Appl. Phys. Lett., 79 (2001) 156.

[3] B. W. D'Andrade and S. R. Forrest, Adv. Mater. 16 (2004) 1585.

[4] S. Liu, J. Huang, Z. Xie, Y. Wang, B. Chen, Thin Solid Films 363 (2000) 294.

[5] J. Thompson, R. I. R. Blyth, M. Anni, G. Gigli, R. Cingolani, Appl. Phys. Lett. 79 (2001) 560.

[6] G. Li, J. Shinar, Appl. Phys. Lett. 83 (2003) 5359.

[7] K. O. Cheon, J. Shinar, Appl. Phys. Lett. 81 (2002) 1738.

[8] C. W. Tang, S. A. VanSlyke, C. H. Chen, J. Appl. Phys. 65 (1989) 3610.

[9] T. Tsuji, S. Naka, H. Okada, H. Onnagawa, Appl. Phys. Lett. 81 (2002) 3329.

[10] B. W. D'Andrade, M. E. Thompson, S. R. Forrest, Adv. Mater.14 (2002) 147.

[11] W. Xie, Z. Wu, S. Liu, S. T. Lee, J. Phys. D: Appl. Phys. 36 (2002) 2331.

[12] H. Choukri, A. Fischer, S. Forget, S. Chénais, M. C. Castex, B. Geffroy, D. Adès, A. Siove, Appl. Phys. Lett. 89 (183513)

[13] S. Grigalevicius, Synth. Met. 156 (2006) 1.

[14] J. F. Morin, M. Leclerc, D. Adès, A. Siove, Macromol. Rapid. Commun. 26 (2005) 761.

[15] D. B. Romero, M. Schaer, M. Leclerc, D. Adès, A. Siove, L. Zuppiroli, Synth. Met. 80 (1996), 271.

[16] V. Boucard, D. Adès, A. Siove, D. Romero, M. Schaer, L. Zuppiroli, Macromolecules 32 (1999) 4729.

[17] A. Fischer, S. Forget, S. Chénais, M-C. Castex, B. Geffroy, D. Adès, A.Siove, J. Phys. D: Appl. Phys. 39 (2006) 917.

[18] Z. Zhi-Lin, J. Xue-Yin, X. Shao-Hong, T. Nagatomo, O. Omoto, J. Phys. D: Appl. Phys. 31 (1998) 32.

[19] H. Aziz and Z. D. Popovic, Appl. Phys. Lett. 80 (2002) 2180.

[20] V. Boucard, T., Benazzi, D. Adès, G. Sauvet, A. Siove, Polymer 38 (1997) 3097.

[21] Commission Internationale de l'Eclairage: "Method of measuring and specifying color rendering properties of light sources", Publ. CIE 13.3-1995

[22] Hernandez-Andres, R.L. Lee Jr, J. Romero, "Calulating correlated color temperatures across the entire gamut of daylight and skylight chromaticities", Applied Optics 38, (1999) 5703-5709.

[23] D. Adès, V. Boucard, E. Cloutet, A. Siove, C. Olivero, M. C. Castex, G. Pichler, J. Appl. Phys. 87 (2000) 7290.

[24] H. Fu, H. Wu, X. Hou, F. Xiao, B. Shao, Synth. Met. 156 (2006) 809.

[25] J. F. Ambrose, L. L. Carpenter, R. F. Nelson, J. Electrochem. Soc. Electrochem. Sci. Techn. 122 (1975) 876.

[26] V. Bulovic, A. Shoustikov, M. A. Baldo, E. Bose, V. G. Kozlov, M. E. Thompson, S. R. Forrest, Chem. Phys. Lett. 287 (1998) 455.

[27] H. Murata, C. D. Meritt, Z. H. Kafafi, IEEE J. Sel. Top. Quantum Electron. 4 (1998) 119.